# 100pT/cm Sensitive MEMS Resonant Magnetometer from a Commercial Accelerometer

Josh Javor[1*], Alexander Stange[2], Corey Pollock[1], Nicholas Fuhr[2] and David J. Bishop[1,2,3,4,5]

1. Department of Mechanical Engineering
2. Division of Materials Science and Engineering
3. Department of Electrical and Computer Engineering
4. Department of Physics
5. Department of Biomedical Engineering

Boston University

Boston, Massachusetts 02215

## Abstract

Magnetic sensing is present in our everyday interactions with consumer electronics, and also demonstrates potential for measurement of extremely weak biomagnetic fields, such as those of the heart and brain. In this work, we leverage the many benefits of the micro-electromechanical systems (MEMS) devices to fabricate a small, low power, inexpensive sensor whose resolution is in the range of weak biomagnetic fields. The sensor works at room temperature, and is suitable for consumer electronics integration. At present, such biomagnetic fields can only be measured by expensive mechanisms such as optical pumping and superconducting quantum interference devices (SQUIDs). Thus, our sensor suggests the opening of a large phase space for medical and consumer applications. The prototype fabrication is achieved by assembling micro-objects, including a permanent micromagnet, onto a post-release commercial MEMS accelerometer. With this system, we demonstrate a room temperature MEMS magnetometer, whose design is only sensitive to gradient magnetic fields and is generally insensitive to the Earth's uniform field. In air, the sensor's response is linear with a resolution of 1.1 nT cm$^{-1}$ and spans over 3 decades of dynamic range to 4.6 µT cm$^{-1}$. In 1 mTorr vacuum with 20 dB magnetic shielding, the sensor achieved 100 pT cm$^{-1}$ resolution at resonance. The theoretical floor of this design is 110 fT cm$^{-1}$ Hz$^{-1/2}$ with a resolution of 13 fT cm$^{-1}$, thus these devices hold promise for both magnetocardiography (MCG) and magnetoencephalography (MEG) applications.

# Introduction

Magnetic sensing spans many scientific applications, from consumer electronics to cutting edge biomagnetic research. Smartphones utilize the Earth's magnetic field for navigation. Automobiles leverage non-contact magnetic sensing to determine position of components, such as in the crank shaft and braking systems. And, more recently, the highest resolution magnetic sensors have been used to measure biomagnetic fields of the brain and heart[1,2]. This large range of applications is accomplished with various sensing mechanisms. Hall Effect sensors dominate the industrial market due to advantages in CMOS-compatible manufacturability and power consumption, but are limited in resolution by the Earth's field at 50 µT (ref. 3). Fields emitted by the brain and heart begin at a million times smaller than Earth's field (approximately 100 pT for heart and 200 fT for brain) and detection must be accomplished by more sophisticated means[1]. Some of the first biomagnetic measurements were conducted by superconducting quantum interference devices (SQUIDs)[4] but have since transitioned to optically pumped, spin-exchange relaxation free (SERF) magnetometers following the development of smaller, chip-scale sensors[5]. Most of these techniques, however, face the tremendous barrier of the Earth's field because they are measuring a uniform field, a magnetic field unchanging in position. For biomagnetic measurements, this requires heavily shielded rooms (typically 60 dB attenuation), averaging or triggering using EKG leads, and large costs (on order of $10k per sensor, QuSpin). To fully realize the clinical capabilities of biomagnetic sensing, arrays including many sensors are needed for biomagnetic mapping, enhancing the impact of cost[6]. In this work, we show that the marriage of a permanent micromagnet and a commercial accelerometer can accomplish both large range (1.1 nT cm$^{-1}$ to 4.6 µT cm$^{-1}$) and high resolution (100 pT cm$^{-1}$) by directly sensing a gradient magnetic field, all with a total cost of goods less than $50. As the first sensor design of its kind, the theoretical floor is 110 fT cm$^{-1}$ Hz$^{-1/2}$, well within the range of biomagnetic field sensing[7]. **Fig. 1** compares our sensor's experiment and theory to existing technology[3,7,8].

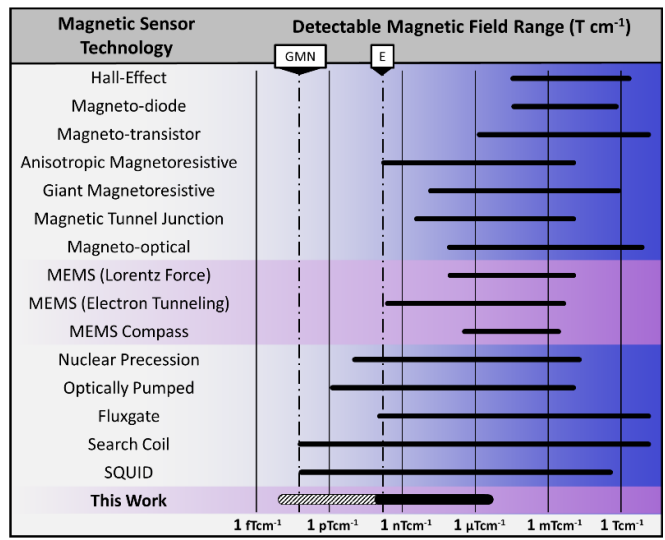

**Fig. 1 Existing Technology.** Sensitivity of various magnetic sensors. Existing sensors measure uniform magnetic field and so are adjusted to measure a gradient field (difference between two sensors separated by 1 cm). E is Earth's gradient field and GMN is geomagnetic gradient noise. This work, an advance for MEMS technology (purple highlight), is shown at the bottom compared to all magnetic sensors[3,7,8]. The dark lines indicate measureable range while the dashed indicate theory. This work is the only intrinsically gradient field sensor.

The key developments enabling the sensor discussed in this work are the highly engineered micro-electromechanical systems (MEMS) accelerometer and the permanent micromagnet. The capacitive accelerometer is the classic success story of the MEMS industry, fulfilling a need in the automotive market for sensitive, low-cost detection for airbag sensors[9]. Such lucrative applications have driven the development of MEMS accelerometers, reaching resolutions of 110 µg Hz$^{-1/2}$ for the ADXL 203 and 20 µg Hz$^{-1/2}$ for the ADXL 354 at costs of $25 and $35 per sensor respectively[10,11]. At the same time, market demand from the hard disk drive industry and others has pushed development of rare-earth permanent magnets. High anisotropy, small size, high remanence, and a large variety of coatings for automotive, medical, and consumer products have led to diverse commercial availability[12]. At first look, the benefits of combining such technologies into a sensor are low power consumption, small size, and low cost. The greatest advantage, however, is that permanent magnets, when constrained from rotating, are only sensitive to forces from gradient fields. So while the Earth's field is large (50 µT), it varies only slightly across the Earth's surface, producing a much smaller gradient field at 500 fT cm$^{-1}$ (ref. 13). Permanent magnets have been integrated into MEMS sensor design before, detecting deflection as a uniform field induces torque, much like in a compass[3,12,13]. The best of these achieved a resolution of 300 pT Hz$^{-1/2}$ at 1 Hz with a large footprint (10 cm$^3$) and electron tunneling feedback control[14]. Several other MEMS magnetic sensors have been designed based on the Lorentz force, achieving the best experimental resolution of 143 nT at 136 kHz (ref. 15,16).

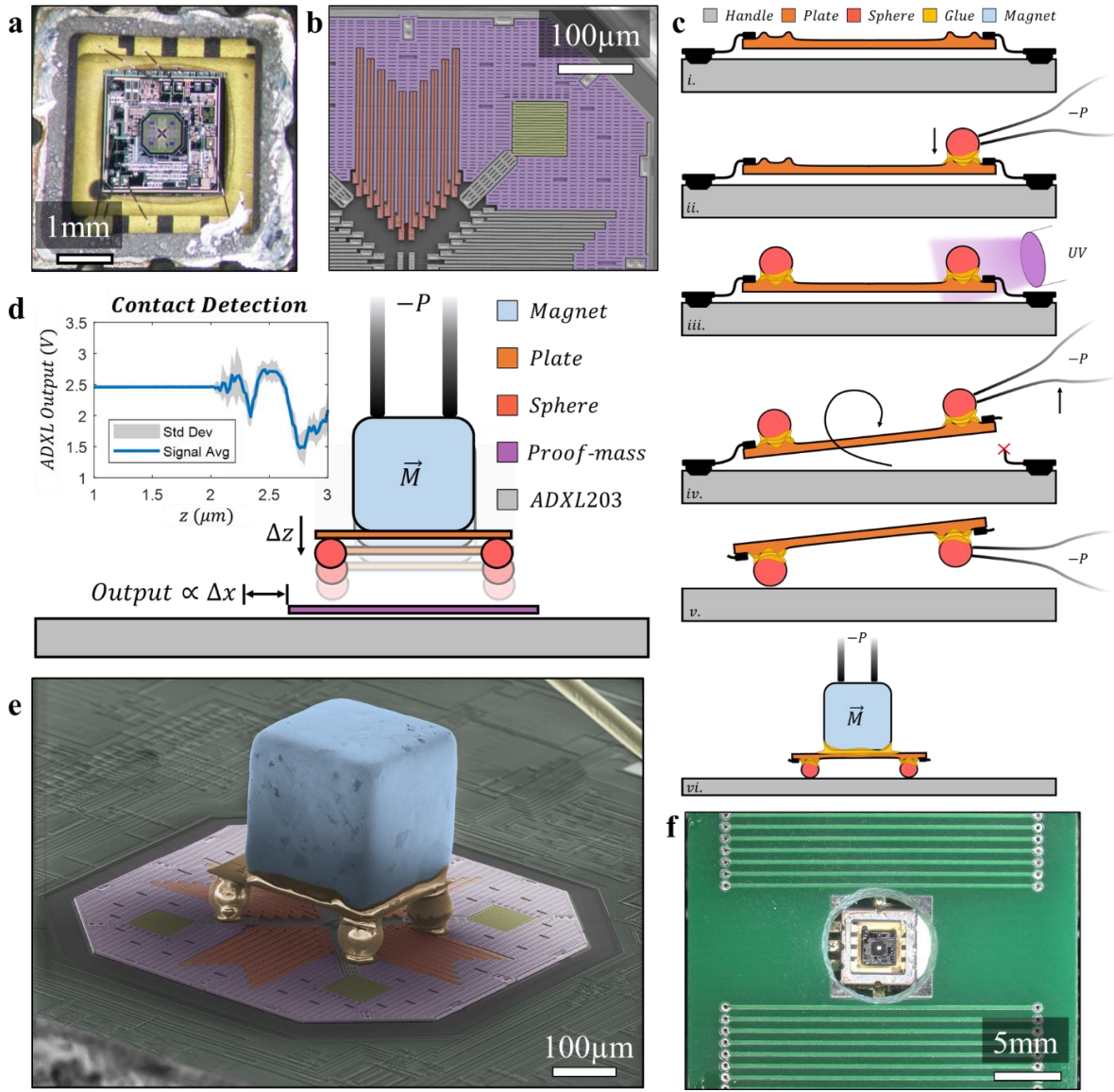

**Fig. 2 Fabrication of Magnetometer. a** Optical image of ADXL 203 accelerometer after hermetic lid removal. Octagonal proof-mass in the center is surrounded by integrated circuitry. **b** Colorized scanning electron microscope (SEM) image of top right quadrant of accelerometer sensor. The proof-mass (purple) is anchored through the springs (yellow) and position is sensed by interdigitated capacitive fingers (red). **c** Schematic of subassembly fabrication. The polysilicon plate (i) is designed in house and fabricated at a foundry. It is mechanically tethered by polysilicon springs (black). A microsphere is manipulated by a micropipette, dipped in UV glue, and placed in a corner of the plate (ii). This is repeated on all corners before the epoxy is cured with UV light (iii). The micropipette is used on one sphere to break the tethers and flip the assembly around to stand like legs under a table (iv – v). A micromagnet is oriented, dipped in UV glue, and then cured in the center of the "table" (vi). **d** Fabrication on post-release MEMS. The subassembly is dipped lightly in UV glue and oriented above the ADXL 203 proof-mass before being lowered carefully until noise is sensed on the output, indicating contact. It is important that the subassembly is attached without interfering with the natural sensing mechanism of the ADXL 203. **e** Colorized SEM image of fully fabricated magnetometer (not the device used in this work). **f** Magnetometer assembled inside printed circuit board (PCB) antiparallel coils for experimental characterization.

# Results
## *Magnetometer Fabrication from MEMS accelerometer*
Our sensor is a marriage of two recently matured technologies: the capacitive accelerometer and permanent micromagnets. The accelerometer is a sensing platform fit for adaptation to other measurands because it inherently senses the position of a movable polysilicon plate. **Fig. 2a** shows the ADXL 203 accelerometer with the hermetic lid removed revealing the silicon die underneath. The octagon in the center is the proof-mass, a quadrant of which is expanded in the **Fig. 2b** false-colored scanning electron microscope (SEM) image. The proof-mass (purple) is a polysilicon plate that can be mechanically coupled to a variety of microscale objects, functionalizing the device for other sensing applications[17].

A custom pick-and-place tool and procedure was developed for assembling of microscale objects on the proof-mass, illustrated in **Fig. 2c and d** (discussed further in Methods). A fully fabricated sensor is shown in **Fig. 2e,** an SEM image. A permanent magnet distorts the SEM image, so the device shown is for illustrative purposes only, where the magnet is completely demagnetized and the UV glue under the spheres overlaps some portions of the spring (compromising sensing ability). Similar fabrication techniques have been used to develop a MEMS Casimir force metrology platform[17], a full hemisphere, tip-tilt micro-mirror[18], and other MEMS at Nokia[19]. **Fig. 2f** shows the sensor fabricated within a custom, printed circuit board (PCB) coil for gradient field characterization.

## *Electrostatic Characterization*
The experimental setup is illustrated in **Fig. 3a**. The sensor can be driven by two mechanisms: electrostatically (purple circuit) and magnetically (green circuit). Further detail is in Methods. The modified ADXL 203 is characterized by electrostatic actuation in **Fig. 3bi** and COMSOL simulation in **Fig. 3b ii-v**. **Fig. 3b i** displays results from a square wave frequency sweep (10 Hz to 3 kHz). The duty cycle is 20% in air and 0.02% in vacuum, maximized to achieve the strongest signal without overdriving at resonance. The root mean square magnetometer output is normalized so that the quality and relative peak magnitude can be compared. Two resonant peaks are shown near 500 Hz and 2.2 kHz and the quality factors (sharpness) of the peaks increase greatly from atmosphere to vacuum, as damping is decreased. The 500 Hz peak is expected as the resonant frequency will decrease from 5.5 kHz when mass is added (see Methods). The quality of the 500 Hz peak is 5 in atmosphere and 4000 in vacuum, demonstrating increased sensitivity at resonance when damping is reduced. **Fig. 3b ii-v** qualitatively illustrate the two modes using the COMSOL Eigenmodes tool. Materials and geometry are input to the model resulting in a calculation and visualization of mode frequency and deformation, respectively. A 3D computer automated design file is generated of the ADXL 203 proof-mass from an SEM image, where the thickness is measured to be 4 μm. The proof-mass is rigidly attached to the magnet-table subassembly, constrained by a roller in the XY plane, and anchored at four points in the center. For this input configuration, a translational mode at 600Hz and a torsional mode at 1.5kHz are found. Errors in the mode frequency are likely due to inaccuracy in model geometry and assumptions of material properties. **Fig. 3b ii** shows a full device view of deformation at the lower frequency mode and **Fig. 3b iii** shows the same mode, cropped and oriented so spring deformation in a quadrant of the proof-mass can be visualized (red is largest deformation, blue is least). The deformation is translational along the X-axis, the direction of magnetization. This is the type of deformation we would expect from a force imposed by a gradient magnetic field (see Methods, **Eq. 5**). Similarly, **Fig. 3b iv-v** show a torsional deformation at the higher frequency mode, where the assembly torques about the center of the x-y plane. This is the deformation we would expect from a uniform magnetic field in x-y plane, but not oriented along magnet's dipole axis (see Methods, **Eq. 7**). Since we are imposing a magnetic field with no uniform component and a constant gradient along the X-axis, we are primarily interested in the effect seen at the translational mode, and we can disregard the higher, torsional mode. Based on this electrostatic characterization, we can simplify our mechanistic understanding of the sensor to a one-dimensional, underdamped harmonic oscillator model (Methods), the free body diagram of which is illustrated in **Fig. 3c**. The collective mass of the subassembly and proof-mass are treated as a rigid body with mass, m. The four springs on the proof-mass are lumped into a single spring constant, k, and damping in air or vacuum modulate the constant, c. An applied force (electrostatic or magnetic) along the X-axis results in a displacement along the same axis. Forces from mechanical and magnetic noise are posited to limit the experimental resolution of the device (see Discussion).

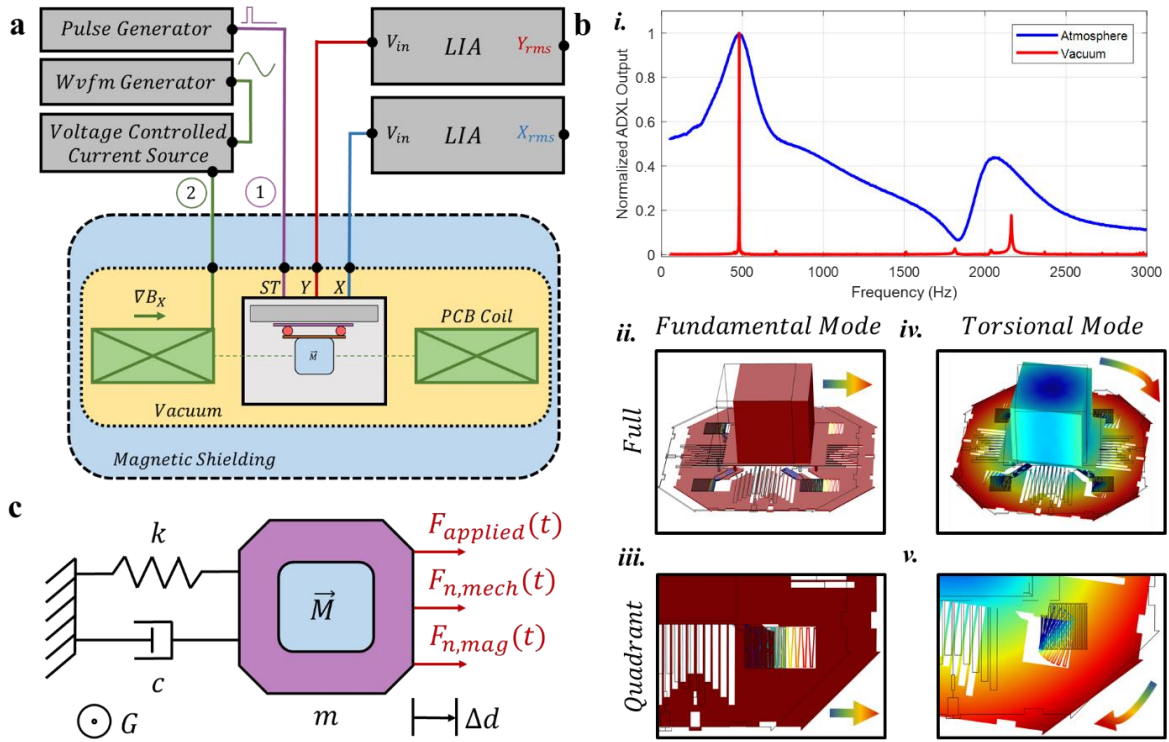

**Fig. 3 Experimental Setup and Electrostatic Characterization. a** Magnetometer on PCB coil assembly is oriented upside down in a chamber with option to pull vacuum (yellow) and apply magnetic shielding (blue). Feedthroughs provide power to the sensor and PCB coil as well as sense the outputs of the functionalized ADXL 203. A pulse generator is used in combination with a built-in self-test (ST) functionality to electrostatically characterize the sensor mechanics in air and in vacuum (purple, 1). A waveform generator is used in combination with a precision current source and the PCB coil wired in antiparallel to magnetically characterize the sensor with gradient fields (green, 2). Both X (blue) and Y (red) outputs are filtered by lock-in amplifiers. **b** Characterization using electrostatic drive described in (**a**) and COMSOL Eigenmode simulation. Square wave frequency sweep (i) reveals two actuation modes and demonstrates increase in quality factor in vacuum. Full view of simulation deflection at first mode depicts translation (ii) and Quadrant view (iii) depicts spring deformation. Full view of simulation deflection at second mode depicts torsion (iv) and Quadrant view (iv) depicts spring and plate deformation. In all color maps, red is largest deformation and blue is least. **c** The simplified free body diagram of the translational mode (top view) resembles a damped harmonic oscillator. Applied force can be driven electrostatically or magnetically. Forces due to mechanical and magnetic noise are also shown (represented as $F_n$).

### *Magnetic Characterization*

The magnetometer's performance is dynamically characterized in three conditions: air (case 1), air with magnetic shielding (case 2), and vacuum with magnetic shielding (case 3). In all cases, the frequency of a gradient field sine wave, symmetric about zero, is swept as the magnetometer output signal is processed by a lock-in amplifier. The magnetometer output voltage is proportional to gradient magnetic field by **Eq 9** (Methods). **Fig. 4a** shows results from case 1, where frequency is swept from 50 Hz to 1.1 kHz and field strength is swept from 4.6 µT cm$^{-1}$ to 1.9 nT cm$^{-1}$. Similar to electrostatic characterization, a low frequency peak is again present near 500 Hz, indicative of the translational mode and displacement along the X-axis. The largest applied field is 4.6 µT cm$^{-1}$, as higher fields result in a clipped output signal at resonance by the ADXL 203 conditioning circuit. The sweeps follow a monotonic pattern, decreasing in signal output as the field magnitude is decreased. At lower field magnitudes, the signal to noise ratio visibly diminishes and is eventually overcome by noise. **Fig. 4b** shows results from case 3, where the applied field is swept in a narrower frequency range on the tip of the high quality peak (478.5 Hz to 480.5 Hz) and in a field range from 3.8 nT cm$^{-1}$ to 76.9 pT cm$^{-1}$. Again, the largest applied field shown is 3.8 nT cm$^{-1}$, above which output signals are clipped by the ADXL 203. The sweeps again follow a monotonic pattern corresponding to field magnitude. The resonance is approximately 479.2 Hz. **Fig. 4c** displays the magnetometer output at resonance with respect to applied gradient field, processed from the sweeps conducted in **Fig. 4a** and **b**. Here, data from sweeps below the experimental resolution of the sensor are included to characterize the experimental noise floor. Data from case 2 (green) is now included, and is largely similar to case 1, except with a lower resolution on the sensing axis. Data from case 1 and case 3 correspond in color to **Fig. 4a** and **b**, respectively. The characteristics from each case are also tabulated in **Table 1**. Circles represent the output from the X-axis (along the magnet's dipole axis), while diamonds are the sensor y-axis output. In all cases, the y-axis is also sensitive to the applied field, but is lower than 14% of the x-output in all cases, indicating good magnet alignment and reduced cross-axis sensitivity. The linear dynamic range of the magnetometer output in fT cm$^{-1}$ is 3.3 decades in case 1 and 1.4 decades in case 3. The dotted lines show a linear least square fit of data above the experimental noise floor, where the sensitivity, $\gamma_{mag}$, is consistently linear and near 1 µVrms (fT cm$^{-1}$)$^{-1}$ in all cases. The black dashed line represents the theoretical noise floor, scaled from the ADXL 203 noise density with optimal lock-in filtering (Methods). The dotted lines are extrapolated to the theoretical floor to show the theoretical resolution in air (80 pT cm$^{-1}$) and vacuum (40 fT cm$^{-1}$). The dash-dot lines are a zero-order, least square fit of data below the experimental resolution, representing an experimental floor in each case. The intersection of the dotted line and dash-dot line are the experimental resolution of the sensor (**Table 1**). It is most noteworthy that x- and y- outputs have the same experimental floor in case 1, and x- and y- outputs reach an experimental floor at the same magnetic field in case 3.
In case 2, the x-output extends lower than both the y-output and case 1 data. These relationships are indications of resolution-limiting noise, expanded further upon in Discussion.

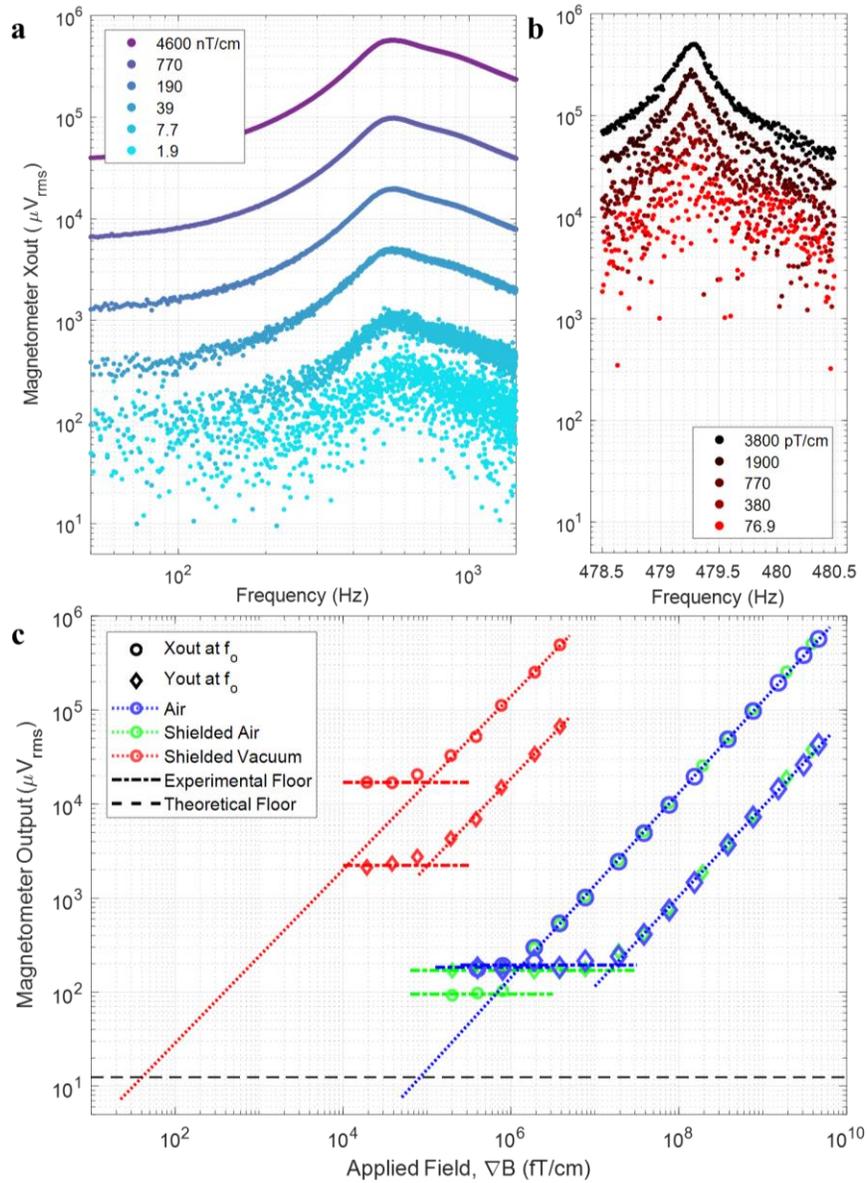

**Fig. 4 Magnetic Sensing Performance. a** Broad sine wave frequency sweep of gradient field in air (around translational mode). Sensor output decreases monotonically with the peak to peak of imposed gradient magnetic field. **b** Narrow sine wave frequency sweep of gradient field in vacuum (around translational mode on tip of high quality peak). Sensor output decreases monotonically with the peak to peak of gradient magnetic field. **c** Magnetometer output at peak of sweeps in (**a**) and (**b**) for air, shielded air, and shielded vacuum conditions. Signals along magnetic direction X-axis (circles) are over an order of magnitude higher than Y-axis (diamonds) indicating good magnet alignment and low cross-axis sensitivity. A linear least squares fit is conducted on data above the experimental floor to determine sensitivity (slope). All sensitivities are nearly 1 μ Vrms (fT cm$^{-1}$)$^{-1}$. Both axes in air reach a floor at the same sensor output, indicating a limitation of mechanical or electrical noise. Both axes in vacuum reach a floor at the same gradient magnetic field strength, indicating a limitation of gradient magnetic noise. In air with shielding, the X-axis reaches a lower sensor output floor than the Y-axis, demonstrating the attenuation of magnetic noise. Based on linear fits, the magnetometer's resolution is 100 pT/cm in vacuum and 1.1 nT cm$^{-1}$ in air. Extrapolating the linear fit further to the theoretical floor, the theoretical resolution of this design configuration is 40 fT cm$^{-1}$ in vacuum and 80 pT cm$^{-1}$ in air, based on ADXL 203 noise density[10] (see Methods).

| Case | Condition | Experimental Resolution (pT/cm) | Theoretical Resolution (pT/cm) | $\gamma_{mag}$, Linear Sensitivity X ($\mu$V / fT/cm) | Cross-Axis Sensitivity (%) | Range ($10^n$ fT/cm) | Experimental Floor X ($\mu$Vrms) | Experimental Floor Y ($\mu$Vrms) |
|---|---|---|---|---|---|---|---|---|
| 1 | Air | 1,050 | 60 | 0.98 | 6.6 | 3.3 | 180 | 190 |
| 2 | Shielded Air | 700 | 60 | 0.99 | 7.5 | 3.6 | 95 | 170 |
| 3 | Shielded Vacuum | 100 | 0.03 | 0.92 | 13.7 | 1.4 | 17,000 | 2,200 |

**Table 1 Sensor Performance Metrics Based on Condition.** Tabulated values are extracted from plot in **Fig. 4c**.

The raw, unprocessed performance of the sensor in air is illustrated in **Fig. 5**, combining some of the performance metrics displayed in **Table 1** (case 1). It is also important to highlight that data in this plot is not taken at resonance, where the signal to noise ratio is far more favorable. Rather, it is operating in a lower frequency regime where many common biomagnetic signals exist. An arbitrary waveform (black, dashed) resembling a magneto-cardiogram is imposed as a gradient field signal at the low frequency of 2.2 Hz. Within a period, the signal is composed of higher frequencies, mostly below 60 Hz. The blue data is the raw output from the magnetic axis and is shown to track the gradient signal very well with no distortion. The red data below the waveform shows the output of the y-axis, showing only mechanical noise and no features from the arbitrary waveform. The inset in the top left is an SEM of the magnetometer showing x- and y-axis direction, where the X-axis is the magnetized direction. Both axes are offset on the plot for ease of visualization. Biomagnetic signatures are typically in the hundreds of pT/cm, and the signal shown here is 20 $\mu$T cm$^{-1}$ peak-to-peak (the smallest feature is a 250 nT cm$^{-1}$ peak indicated at 0.7 s). While this is several orders of magnitude away, the Discussion expands on why this is a promising platform for such measurements in the future.

## Discussion

### *Fabrication Error and Throughput*

In **Fig. 2**, we show that a commercially available accelerometer can be functionalized with a micromagnet using a custom pick-and-place procedure[17,18]. Previous prototypical works combining permanent magnets with MEMS structures did not employ commercially available platforms or such position-sensitive fabrication techniques[3,14-16]. While the fabrication method presented here is currently low-throughput and useful mainly for prototypical design, industrial scale fabrication techniques exist to accomplish similar tasks, such as pick-and-place or flip-chip-bonding[20,21], and could be used to fabricate the sensor pre-release. It is also noteworthy that several asymmetries result from such a manual fabrication process that could limit the resolution of the sensor. Some of these include a displaced center of mass (anisotropy of cube magnet geometry, centering of magnet on table, table on proof-mass), variable sphere size and area of contact, magnet orientation, magnetization direction, and contamination of sensor from opening the sealed package. With the added weight of the micromagnet, such asymmetries may manifest themselves by pulling the proof-mass out of plane with respect to the capacitive fingers, which are designed to detect displacement in-plane only. Any rotational assembly error or uniform field (such as the Earth's field) may create an offset torque of the proof-mass, creating greater asymmetry in the springs and mechanical motion. Altogether, we show that the fabrication technique has minimized these errors and suggest that existing large-scale systems could reduce them even further.

## Exclusive Sensitivity to Gradient Magnetic Fields

The mechanical modes of our magnetometer are characterized by electrostatic actuation in **Fig. 3b i** and by mode simulation in **Fig. 3b ii-v**. We argue that the fabricated sensor is only sensitive to gradient magnetic fields, which impose a force along the dipole axis of the magnet and result in a translational deformation (Methods). Sensitivity to uniform fields would result in a torque of the magnet. This is supported by the simulated mode deformation in **Fig. 3b ii**, where we show that the fundamental mode is a translational deformation along the dipole axis of the magnet. Thus any information from this mode will only be from gradient magnetic fields. We also explain the prediction of the 500 Hz fundamental mode from a simplified free body diagram in **Fig. 3c** and **Eq 3** (Methods). Furthermore, we show there is a separate, higher mode of torsional deformation (**Fig. 3 iv**) about the center of the magnet that would correspond to a uniform field. This shows that information at the fundamental mode is not directly affected by uniform fields. The magnetic characterization of the device, then, is centered around the fundamental mode.

## Resolution-Limiting Noise

The magnetic characterization in **Fig 3c** and **Table 1** reveal differing resolutions in each case and on each output axis of the fabricated sensor. Analyzed together, the experimental noise floors are suggestive of the type of noise that limits the resolution. In air (case 1), the experimental floors of both x- and y- outputs coincide at the same sensor voltage, indicating that both axes are subject to a common noise. Since the noise floor is independent of magnet orientation, it cannot be due to gradient magnetic field noise, which would predominantly actuate the magnet's central axis. Mechanical noise may result from the experimental set up or air pressure fluctuations in the ambient environment (as the sensor's lid is removed). Any asymmetries from fabrication discussed earlier could result in common electrical noise on the output from out-of-plane capacitive fingers. The directionality detection of the ADXL 203 is designed to modulate each axis differently, and out-of-plane deformation could enhance cross-talk of these signals. Furthermore, the device is unshielded in case 1 and an offset torque from the Earth's uniform field could add to the asymmetry. When shielded in case 2, the X-axis reaches a lower resolution (from 1.1 nT cm$^{-1}$ in air to 700 pT cm$^{-1}$ in shield), but the y-axis remains the same. The effect of the shielding may be damping of mechanical noise (stabilizing the motion of the magnet) or attenuating any offset torque from the Earth's field that could be adding to asymmetry. It is unlikely that uniform geomagnetic noise plays a role here as geomagnetic noise is far lower at 100 pT (ref. 3).

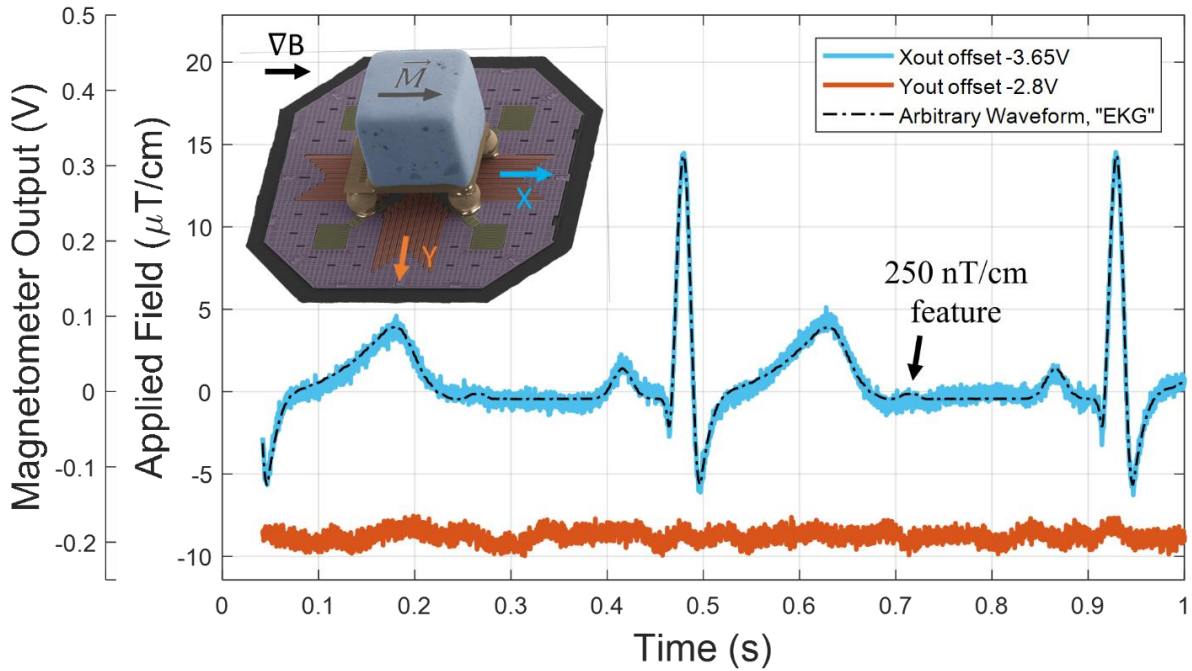

**Fig. 5 Magnetic Sensing Demonstration.** Raw output of magnetic sensor in air in response to an arbitrary waveform resembling an electrocardiogram (EKG) at 2.2 Hz and 20 µT cm$^{-1}$ peak-to-peak, imposed along the X-axis by a PCB coil. The sensor output is displayed both in voltage and gradient magnetic field. Both X and Y outputs are offset for ease of visualization. The X output tracks the imposed field very well, while the Y output does not resolve any of the features in the magnetic signal. The inset (top left) illustrates the magnet alignment with respect to the sensitive X-axis and the insensitive Y-axis.

Results from the shielded vacuum condition (case 3) reveal a common experimental floor of both sensor axes at the same applied field, rather than the same sensor output voltage. This indicates that the limiting noise source is different than that in case 1 and case 2. The y-axis motion was still able to be detected at much lower displacements than the X-axis, suggesting that the limiting noise must be a gradient magnetic noise. Possible sources are either geomagnetic or the gradient coil driving system. Geomagnetic gradient noise has been reported to be much lower at 500 fT cm$^{-1}$ (ref. 13), however measurements were conducted during daytime hours in a major city. Therefore, gradient noise from the environment may be larger than this report, but are likely not the limiting noise. The voltage controlled current source (CS580) is specified to have superb output noise characteristics (60 fA Hz$^{-1/2}$ in the configuration at resolution measurement). However, the instrument is some distance away from the PCB coils, the wire is carried next to all other sensor leads with relatively small drive currents (<100nA drive, 11mA power), and connections are made at vacuum chamber feedthroughs. This may make the drive signal vulnerable to pick-up or cross talk, which is amplified and superimposed onto the magnetic driving force, thus limiting the sensor resolution.

## *Current and Future Applications*

The sensor is characterized in three conditions (**Table 1**) to demonstrate its performance in the context of various applications. Most applications exist in ambient conditions and don't require the enhanced performance that vacuum and shielding provide. For example, dipole sources, such as planetary magnetic fields or ferromagnetic objects, have gradient signatures that are difficult to measure with a uniform field sensor alone. Among these applications, our magnetometer offers the key advantage of directly sensing the gradient field, rather than the difference between two uniform field sensors. Moreover, it is capable of doing all this with a small size, low power, low cost, and at room temperature. Finally, the realization of our theoretical resolution in air (80 pT cm$^{-1}$) would offer the unique ability to sense biomagnetic fields in ambient conditions, an idea that is attractive for wearable sensing (such as signals illustrated in **Fig. 5**).

For the most sensitive applications, vacuum or shielding can be applied. Vacuum can be pulled on the resonant sensor as shown in **Fig. 3b i** and **4b** to increase the quality and resolution of the fundamental mode. At scale, vacuum packaging is a solved problem for MEMS packaging[21], which allows for the benefits of enhanced resolution at a small scale. Resonant mode operation is typically a design tradeoff, limiting a sensor to a specific, narrow band of frequencies around resonance. However, a resonant sensing mechanism does not necessarily impede a resonant sensor from identifying features at other frequencies of interest. MEMS actuators with flux guides have been designed to modulate an arbitrary signal so it can be measured at the resonant frequency of a sensor[22]. Others have leveraged a nonlinear spring stiffness during cyclic resonant motion, ultimately reporting a shift in resonant frequency instead of oscillation amplitude[23]. Shielding becomes useful when the target of measurement can also fit inside the shield. For this, shielded rooms with 60 dB attenuation are common and often used for biogmagnetic measurements. We report the effect of just 20 dB magnetic attenuation, which demonstrates the potential for enhanced resolution with greater shielding. Finally, the combination of improved vacuum, shielding, and environment could realize a theoretical resolution (13 fT cm$^{-1}$) directly on par with SQUIDs and optically pumped, atomic magnetometers. Such a sensor is disruptive in cost, size, and its gradient sensing mechanism, transforming the approach to the most sensitive applications such as biomagnetic fields.

# Materials and Methods
## *MEMS Accelerometer*
**Fig 2b** shows an SEM image of a quadrant of the ADXL 203 from Analog Devices. The spring (yellow) is nearly symmetric on both X and Y axes. Displacement is sensed via capacitive fingers (red) in a differential configuration. We chose the ADXL 203 for its intrinsically low noise density (110 µG Hz$^{-1/2}$), linearized sensitivity (1 V G$^{-1}$), wide range (up to 10$^4$ G with optimized filtering), and accessible proof-mass[10]. From experimental observation, the two-axis accelerometer was found to have a resonant frequency of 5.5 kHz, a spring constant of 1 N m$^{-1}$, a Q in air of 10, and Q in vacuum of 10,000. The maximum sensing range in one direction is 25 nm, giving a sensitivity of 10 nm V$^{-1}$. This means the device has a noise density of 1 pm Hz$^{-1/2}$, or 1 pN Hz$^{-1/2}$.

## *Permanent Micromagnets*
Permanent micromagnets are a sintered mix of rare-earth element powders and are typically coated for protection or passivation[12]. The micromagnets in this work are cubes and magnetized to N52 grade (SM Magnetics Co.). The smallest commercially available cube magnet of 250 µm side length is chosen to minimize gravitational forces. The powders consist of neodymium, iron, and boron. The standard coating of nickel, copper, and nickel is used to avoid degradation.

## *Magnetometer Fabrication*
The magnetometer is fabricated in two stages. First, a subassembly is made that resembles a magnet on a table (**Fig. 2c**). Second, the subassembly is attached to the post-release MEMS (**Fig. 2d**). The separate subassembly comprises spheres, a polysilicon plate, and a micromagnet. A custom pick-and-place system is used to manipulate and assemble these micro-objects. Vacuum (typically -2psi) is pulled on a glass pipette with the orifice in contact with the object of interest. A micromanipulator on a probe station (Cascade Microtech EPS150FA) and a 3D printed part are used to direct the motion of the pipette in 3 dimensions. A straight pipette (WPI, 30 µm aperture) is used to manipulate microspheres and a custom, 45° angled pipette (Clunbury Scientific, 135 µm aperture) is used with the cube magnet. Borosilicate glass microspheres (Cospheric) of about 65 µm are used as they are sufficiently larger than the pipette, but still small enough to minimize contact with the MEMS proof-mass later. The plate is designed in-house and manufactured by the MEMSCAP foundry process, PolyMUMPs (**Fig. 2c i**). Mechanical tethers of polysilicon are attached to both the plate and the silicon handle so that the plate is suspended when a sacrificial layer of oxide is removed underneath by etching with hydrofluoric acid.

Spheres are assembled on the plate to form a "table," which the magnet sits on. The spheres minimize contact surface area between the micro-objects and the sensitive proof-mass, allowing for repeatable, robust assembly. To assemble a sphere on the plate, vacuum is pulled while in contact with a sphere, which is then wetted on the bottom side with UV glue (Norland Adhesives, NOA 81). The sphere is then positioned above and lowered onto a corner of a plate (**Fig 2c ii**), where it is partially cured by

UV light at the manufacturer recommended wavelength of 365nm (Dymax BlueWave) for 15 seconds (**Fig 2c iii**). This is repeated until all four corners of the plate contain spheres. Next, the vacuum pipette is put in contact with one sphere to form a "ball in socket" joint. The pipette is lifted to break the mechanical tethers holding the plate (**Fig 2c iv**). The newly assembled "table" is turned to sit on its legs (**Fig 2c v**). Next, a micromagnet is oriented on a vertical glass slide by an external magnet some distance behind the slide. The larger, angled pipette is brought into contact with the top face of the magnet, which is not one of the poles. The external magnet is removed, leaving magnet on the end of the pipette, held by vacuum. The bottom of the magnet is then dipped in UV glue, aligned on the standing table, and radiated with UV light.

Now that the subassembly is a single rigid structure, the large, angled pipette can be used to attach it to the post-release MEMS. The sensor lid is removed and power is supplied so that the noise on the X and Y outputs can be monitored. The noise is very low normally, but spikes when contact is made by the pick-and-place system, presumably from vibrations in the pipette (**Fig. 2d** inset). The subassembly structure is lifted up and aligned over the center of the proof-mass under a brightfield microscope. The spheres are in predefined locations on the plate so that they make contact with a strip of the proof-mass between the capacitive fingers and the spring. The spheres on the bottom of the plate are dipped very lightly in UV glue, before the structure is carefully lowered toward the proof-mass (**Fig. 2d**). A spike in the accelerometer output signal is used to detect contact between the subassembly and the proof-mass (**Fig. 2d** inset). The spheres minimize contact with the proof-mass so that epoxy does not wick through the release holes patterned on the proof-mass, in which case the device would be rendered insensitive. The structure is then radiated with UV light for 15 seconds before cutting vacuum to the pipette and lifting off. The entire structure is then baked upside-down (to avoid unintended gluing) at 60C overnight to form a full cure (below manufacturer-recommended maximum operating temperature of magnet).

## *Experimental Setup and Measurement*

The experimental setup comprises a custom PCB coil, custom vacuum chamber, MuMetal shield (Magnetic Shields Corp.), and an instrument drive system (**Fig. 3a**). The ADXL 203 is surface-mounted on a custom printed circuit board (PCB), on which a PCB coil for magnetic characterization is also attached (**Fig. 2f**). The PCB coil consists of two layers separated by the 1.6 mm thick PCB. The top copper traces can be seen in the image and the bottom traces are only different where the end of each line connects it to the next winding in sequence. The 0.1 mm vias, spaced 1 mm apart, connect the two layers to form a 10 turn coil on either side of the sensor. The PCB coil is rigidly connected to the PCB board to reduce mechanical noise in the output. The coil pair central axis is aligned with the micromagnet's dipole axis (along the accelerometer X-axis) for magnetic drive. Force on the magnet is proportional to the gradient of the magnetic field. By wiring the PCB coil in antiparallel fashion, the resulting magnetic field has a constant slope relative to position across the sensor, and thus a gradient magnetic field. The uniform field, then, is zero at the center of the coils, where the micromagnet is positioned. The sensor-coil assembly is fit into a dual inline pin (DIP) socket within a vacuum chamber built using standard parts (Kurt J. Lesker). For experimental results, the chamber is either at atmospheric pressure or in vacuum (1 mTorr). The sensor is positioned upside down to keep the proof-mass free from contacting the substrate underneath in the event of off-axis fields. The chamber is held between vibration isolating pads on a two-axis vertical stage, allowing for the chamber to be moved in and out of a MuMetal shield, which attenuates 20 dB of imposing fields (Magnetic Shields Corp). This entire assembly is built on a passive hydraulic vibration isolation table. There are two driving schemes of the magnetic sensor shown in **Fig. 3a** drive schematic: Electrostatic actuation and Magnetic actuation. Electrostatic actuation (circuit 1, purple) leverages the capacitively driven self-test functionality of the accelerometer. Originally designed to test whether the accelerometer is in normal working condition, this function may also be used to drive the sensor to arbitrary positions in one quadrant of the actuation range using pulse width modulation (PWM)[24]. In this work, self-test is used to non-magnetically actuate the magnetometer over a range of frequencies, characterizing the mechanics of the device after micro-objects are attached. The self-test pin is driven by a precision pulse generator (SRS DG645), as short duration pulses (<200 ns) are required in vacuum to avoid over-driving the MEMS at resonance. The pulse is 0 to 3 V and duty cycle is 20% in air and 0.02% in vacuum. Magnetic actuation (circuit 2, green) is achieved by driving low noise, small (>100 fA) currents through the PCB coil using a voltage-controlled current source (SRS CS580) and waveform generator (Agilent 33210A). The PCB coil's gradient field is linear with drive current (760 µA (µT cm$^{-1}$)$^{-1}$) and unchanging (< 3%) with frequency in the actuation range (DC to 1 kHz). In both drive schemes, the sensor output is only filtered using a lock-in amplifier (SR830). The reference signal is a 50% duty cycle square wave from either the pulse generator or the waveform generator. The equivalent noise bandwidth (ENBW) when using a 24 dB oct$^{-1}$ roll-off and 300 ms time constant is equal to 0.26 Hz. The minimum ENBW for the most sensitive measurements is 0.008 Hz. Resolution at a given frequency is calculated from noise density, $\rho$, by **Eq 1** (ref. 25), and is elaborated on further in **Eq 10**.

$$Resolution = \rho * \sqrt{ENBW * 1.6} \quad (1)$$

## *Measurement Theory*

### Governing Mechanics

A simplified, one-dimensional mechanical model is shown in **Fig. 3c**. The magnetometer behaves as an underdamped harmonic oscillator (**Eq 2**), with settling times of 20 ms in air and 3 s in vacuum (within 2% of final value).

$$m\ddot{d} + c\dot{d} + kd = F_{applied}(t) + F_{n,mech}(t) + F_{n,mag}(t) \quad (2)$$

The system may be thought of as one-dimensional because the ADXL 203 has low cross-axis sensitivity (1.5%) between X and Y[10] and out-of-plane forces are minimized by centering the magnet on the proof-mass using the table subassembly. The magnet, table, and proof-mass are all considered one rigid body. Together, their mass is found to be 160 µg from the relationship between resonant frequency and mass shown in **Eq 3**. The resonant frequency is found from an electrostatic frequency sweep, shown in **Fig. 3bi**. This explains the effect of a decreased resonant frequency when mass is added. The spring constant, k, is estimated to be near 1 N m$^{-1}$ from SEM images of the spring and a COMSOL model assuming polysilicon material.

$$m = \frac{k}{(2\pi f_0)^2} \quad (3)$$

The constant, c, in **Eq 2** represents damping. Vacuum decreases damping and is shown to increase the quality factor in **Fig. 3b i**. At resonance, this increases the amplitude of oscillation, $d_{f0}$. For a constant force at resonance, $F_{f0}$, the amplitude increases proportional to the quality factor, $Q$ as shown in **Eq 4** (ref.9).

$$d_{f0} = Q \frac{F_{f0}}{m} \quad (4)$$

### Forcing and Magnetics

The applied force, $F_{applied}$, is proportional to the gradient field[26] generated by the PCB coil (**Eq 5**) as described in the previous section. The micromagnet has a moment, $\vec{M}$, of 15 µJ T$^{-1}$ calculated from experimental data and confirmed by simulation. The permanent magnet is approximated as a dipole in **Eq 6** and data is fit by cubic function[26]. Experimental magnetic field, B, is gathered from a hall sensor along the central axis of the magnet, r. The constants, including magnetic permeability of free space, $M_0$, are condensed to $\alpha$, and the moment of the magnet, M, is extracted. This is confirmed by a simulation using Finite Element Methods Magnetic (FEMM) software and a 250 μm cylindrical magnet of N52 grade. Again magnetic field, B, is collected at various distances along the central axis, r, and fit by cubic function to extract the moment, M.

$$F_{applied}(t) = M \cdot \nabla B(t) \quad (5)$$

$$B(r) = \frac{2M_0 M}{4\pi} \frac{1}{r^3} = \frac{\alpha M}{r^3}, \quad (6)$$

The main sources of noise are mechanical, $F_{n,mech}$, and gradient magnetic, $F_{n,mag}$, but these are insignificant for large fields and so are analyzed in Discussion as resolution-limiting terms. The effect of gravity is ignored since the magnetometer is held upside down and any gravitational forces would be outside of the sensing plane. This simplified model is only relevant to the fundamental mode in **Fig 3b ii**, as the mode described in **Fig. 3b iv** is deforming in two dimensions. This torsional deformation could be actuated by a uniform magnetic field, much like a compass. The relationship between a uniform field, B, and the magnet with moment, M, would be a torque, T, as is shown in **Eq 7** (ref. 26). The ADXL 203 is designed to sense motion in either the x or y directions, however. And so in this case, the sensor would not have a meaningful output.

$$T = M \times B \quad (7)$$

Sensor Transduction

The ADXL 203 directly measures a differential capacitance, which is inversely proportional to a displacement of the proof-mass (in either X or Y directions, see **Fig. 2b**). The sensor is linear in the measurement range and so its signal output, $S$, is related to displacement, $d$, by a proportionality constant, $\gamma$, in **Eq 8**. In the linear regime of the springs, Hooke's Law relates displacement to a force. **Eq 5** earlier shows that $\nabla B$ is proportional to a force, $F_{applied}$. Once functionalized with a micromagnet, the sensor output, now $S_{mag}$, is then proportional to the applied $\nabla B$ by the sensitivity, $\gamma_{mag}$ (**Eq 9**). The sensitivity of the magnetometer is 1 μV (fT cm$^{-1}$)$^{-1}$ from experimental measurement in **Fig 4c** and tabulated in **Table 1**.

$$S = \gamma d \quad (8)$$

$$S_{mag} = \gamma_{mag} \nabla B \quad (9)$$

Noise and Theoretical Resolution

Using the understanding of mechanics and effect of magnetic fields on the sensor output, the noise floor can be calculated. The theoretical resolution is assumed to be limited by the noise floor of the sensor. The ADXL data sheet reports a noise density, $\rho$, of 110 μV Hz$^{-1/2}$ (ref. 10). This accelerometer noise density can be scaled by the magnetic sensitivity, $\gamma_{mag}$, in **Eq 10** to find magnetic noise density, $\rho_{mag}$. This yields a magnetic noise density of 110 fT cm$^{-1}$ Hz$^{-1/2}$. Finally, **Eq 1** earlier related noise density to sensor resolution when using a lock-in amplifier. This experimental setup, then, would reach a theoretical best measurement of 13 fT cm$^{-1}$ in 1 mTorr vacuum and at room temperature.

$$\rho_{mag} = \gamma_{mag} \rho \quad (10)$$

## Conclusion

The purpose of this work was threefold: 1) to build a magnetic sensor that is only sensitive to gradient magnetic fields, 2) to demonstrate the wide field and frequency space of this new class of magnetic sensors, and 3) to achieve 1 and 2 in a small, low-cost, and commercially available platform. We have demonstrated the performance of a gradient magnetometer by achieving a resolution of 100 pT cm$^{-1}$ in shielded vacuum, and a range spanning over 3 decades in ambient conditions (1.1 nT cm$^{-1}$ to 4.6 µT cm$^{-1}$). Compared to existing designs of magnetic MEMS resonant sensors, our resolution surpasses the best found in literature[14] by a factor of three experimentally and by over a factor of fifteen theoretically (110 fT cm$^{-1}$ Hz$^{-1/2}$). More sensitive accelerometers, such as the ADXL 354, would theoretically be able to improve this sensitivity by a factor of ten[11]. We have achieved all of this on a small, versatile platform, which is easily integratable into consumer technology integration. This new technology has potential to revolutionize magnetic sensing while also offering many advantages to other fields such as navigation, communication, and biomagnetic field mapping.

## Acknowledgements

We would like to thank Pablo del Corro, Lawrence Barret, and Jeremy Reeves for helpful consultations regarding state-of-the-art magnetometry, theory, and fabrication. We would also like to thank Professor Anna Swan for the customization of her probe station for sensitive fabrication. This work was supported by the NSF CELL-MET ERC award no. 1647837 and a SONY Faculty Innovation Award.

## Conflict of Interests

The authors declare that they have no conflict of interest.

## Author Contributions

The device and experiments were conceived by D.J.B. and J.J. The fabrication was done by J.J. with assistance from A.S. All data were collected by J.J. and interpreted and analyzed by J.J., A.S., C.P., N.F., and D.J.B. The manuscript was written by J.J. with input from A.S. and edited by all authors.